\documentclass[aps,prd,twocolumn,superscriptaddress,preprintnumbers,floatfix,nofootinbib,notitlepage,showkeys]{revtex4-1}
\usepackage{multirow}
\usepackage{siunitx}

\usepackage{graphicx,times}
\usepackage{latexsym}
\usepackage{mathtools}

\usepackage{amsmath,amssymb,amsbsy,amsfonts}
\usepackage{array}
\usepackage{bm}
\usepackage{graphics}
\usepackage{mathrsfs}
\usepackage{xcolor}
\usepackage{cancel}
\usepackage[normalem]{ulem}

\usepackage{hyperref}
\usepackage[capitalise]{cleveref}

\usepackage{xcolor}
\usepackage{makecell}
\newcommand\TopRule{\Xhline{0.08em}}

\newcommand\MidRule{\Xhline{0.03em}}
\newcommand\BotRule{\Xhline{0.08em}}

\setcellgapes{2pt}

\newcommand{\tD}{{\widetilde{\Delta}}}

\newcommand{\psib}{{\overline{\psi}}}

\renewcommand\>{\rangle}
\newcommand\<{\langle}

\newcommand\StateIn[2]{{|{#1};{#2}\>}}
\newcommand\StateOut[2]{{\<{#1};{#2}|}}

\begin{document}

\title{\texorpdfstring{Sub-leading conformal dimensions at the O(4) Wilson-Fisher fixed point}{Sub-leading conformal dimensions at the O(4) Wilson-Fisher fixed point}}
\author{Debasish Banerjee}
\affiliation{Saha Institute of Nuclear Physics, Bidhan Nagar, Kolkata, West Bengal 700064, India}
\author{Shailesh Chandrasekharan}
\affiliation{Department of Physics, Box 90305, Duke University, Durham, NC 27708, USA}

\begin{abstract}
In this work we focus on computing the conformal dimensions $D(j_L,j_R)$ of local fields that transform in an irreducible representation of $SU(2) \times SU(2)$ labeled with $(j_L,j_R)$ at the $O(4)$ Wilson-Fisher fixed point using the Monte Carlo method. In the large charge expansion, among the sectors with a fixed large value of $j = {\rm max}(j_L,j_R)$, the leading sector has $|j_L-j_R| = 0$ and the sub-leading one has $|j_L-j_R| = 1$. Since Monte Carlo calculations at large $j$ become challenging in the traditional lattice formulation of the $O(4)$ model, a qubit regularized $O(4)$ lattice model was used recently to compute $D(j,j)$. Here we extend those calculations to the sub-leading sector. Our Monte Carlo results up to $j=20$ fit well to the form $D(j,j-1)-D(j) \sim \lambda_{1/2}/\sqrt{j} + \lambda_1/j + \lambda_{3/2}/j^{3/2}$, consistent with recent predictions of the large charge expansion. Taking into account systematic effects in our fitting procedures we estimate the two leading coefficients to be $\lambda_{1/2}=2.08(5)$, $\lambda_1=2.2(3)$.
\end{abstract}

\maketitle

\section{Introduction}

There has been a resurrection of interest in conformal field theories in recent years, especially due to the success of the bootstrap approach in certain problems \cite{Rychkov:2016iqz,Simmons-Duffin:2016gjk}. It has also become clear that conformal field theories simplify in sectors with either large spin \cite{Komargodski:2012ek} or large global charge
\cite{Hellerman:2015nra}. Due to these developments the field has seen a renaissance with several new results over the past few years \cite{Kaviraj:2015cxa,Kaviraj:2015xsa,Alday:2016njk,Gopakumar:2016wkt,Dey:2017fab,Caron-Huot:2017vep,Hellerman:2017sur,Hellerman:2017veg,Jafferis:2017zna,Alvarez-Gaume:2019biu,Orlando:2019skh}. A recent review of the above progress can be found in Ref.\cite{Poland:2018epd,Dondi:2021buw}. 

An interesting quantity in a conformal field theory is the conformal dimension of local fields that transform according to some representation of the symmetries of the theory. In this work we focus on global symmetries and study CFTs that emerge in three-dimensional $O(N)$ models at the Wilson-Fisher fixed point. Recent work \cite{Hellerman:2015nra} showed that in these theories the conformal dimension $D(Q)$ of local fields that transform under the representation with charge $Q$ satisfy a large charge expansion of the form 
\begin{align}
D(Q) = \sqrt{\frac{Q^3}{4\pi}}
\Bigg(c_{3/2}+\frac{4\pi}{Q}c_{1/2} + {\cal O}\Big(1/Q^2\Big) \Bigg)  + c_0,
\end{align}
where $c_{3/2}$ and $c_{1/2}$ are low-energy constants that need to be determined non-perturbatively, while $c_0 \approx -0.094$ can be determined analytically \cite{PhysRevD.94.085013,daleFuente18}. Non-perturbative Monte Carlo calculations confirming these predictions for the case of the $O(2)$ model, where the global charges $Q$ are represented by integers, have also been performed, and it was discovered that in this case $c_{3/2} = 1.195(10)$ and $c_{1/2} = 0.075(10)$ \cite{PhysRevLett.120.061603}. These calculations were later extended to the $O(4)$ model where the local fields transform in some representation of the $O(4)$ symmetry. We can classify them according to the irreducible representations of $SU_L(2)\times SU_R(2)$ labeled with charges $(j_L,j_R)$ where $j_L,j_R=0,1/2,1,3/2,...$. In this case it is natural to define the charge as $Q = 2j$ where $j={\rm max}(j_L,j_R)$. In the large charge expansion, the leading sector is given by $|j_L-j_R|=0$ sector, for which Monte Carlo calculations give $c_{3/2}=1.068(4)$ and $c_{1/2}=0.083(3)$ \cite{Banerjee:2019jpw}. It was shown that the large charge expansion predicts the conformal dimensions, even at $Q=1$ within a few percent.

How well does the large charge expansion do in predicting the conformal dimensions in the sub-leading sector, where $|j_L- j_R|=1$? This is the question that motivates our research in this work. A natural quantity to measure in this case is 
$\tD(j) = D(j,j-1)-D(j,j)$, which has the expansion of the form 
\begin{align}
\tD(j) &= \lambda_0 + \frac{\lambda_{1/2}}{j^{1/2}} + \frac{\lambda_1}{j} + \frac{\lambda_{3/2}}{j^{3/2}} + {\cal O}\Big(\frac{1}{j^2}\Big),
\label{eq:lqpred}
\end{align}
valid for large values of $j$. In this expansion, the coefficients of fractional powers of $j$ like $\lambda_{1/2}$ and $\lambda_{3/2}$ are Wilson coefficients that cannot in principle be determined within the large charge effective field theory \cite{Gaume:2020bmp}. Their origin is essentially classical with quantum corrections. On the other hand the coefficients of integer powers of $j$ like
$\lambda_0$ and $\lambda_1$ arise from purely quantum mechanical effects and can in principle be calculable analytically, similar to $c_0$ introduced earlier. It is conjectured that $\lambda_0=0$ since it is difficult to imagine a calculation that would distinguish between $D(j,j)$ and $D(j,j-1)$ in the large $j$ limit \cite{Err1}. However, the spin of the leading conformal field in the $(j,j)$ sector is different from the $(j,j-1)$ sector. Such differences were first observed in \cite{Alvarez-Gaume:2016vff} and later clarified in the context of $O(4)$ in \cite{Hellerman:2017efx,Antipin:2020abu}. For this reason $\lambda_0 \neq 0$ may still be possible. On the other hand $\lambda_1$ is most likely non-zero but remains undetermined until now \cite{Com1}.

In this work we design a new Monte Carlo method to compute $\tD(j)$ in order to explore the validity of \cref{eq:lqpred}. We present results in the range $1 \leq j \leq 20$. Within this range we study if the conjecture that $\lambda_0=0$ is consistent with our results and estimate the other three unknown coefficients.

\section{Qubit Regularized O(4) model}

In order to construct a Monte Carlo method to compute $\tD(j)$ it is useful to understand how the $SU_L(2)\times SU_R(2)$ symmetry is manifest in our lattice model, which is the same as the one used in Ref.~\cite{Banerjee:2019jpw}. For this purpose it is helpful to view our model as strongly coupled lattice Quantum Electrodynamics (QED) constructed with staggered fermions \cite{Cecile:2007dv}. When gauge fields are integrated out exactly, the microscopic degrees of freedom are made up of bosons with fermionic constituents. These bosons naturally have a built-in hardcore interaction, and hence the Hilbert space on each lattice site is finite dimensional. Such bosonic lattice field models with a finite dimensional Hilbert space, that reproduce a continuum quantum field theory, can be referred to as a qubit regularized model of the continuum quantum field theory \cite{Singh:2019uwd}. Other examples of qubit regularized models for studying continuum quantum field theories with $O(N)$ symmetries, have been constructed recently \cite{Singh:2019jog,Bhattacharya:2020gpm}. 

The lattice action of our qubit regularized $O(4)$ model can be written using four Grassmann valued lattice fields $\psi_{1,k},\psi_{2,k},\bar{\psi}_{1,k},\bar{\psi}_{2,k}$ at each lattice site $k \equiv ({\mathbf{r},\tau})$ on a cubic lattice, where we distinguish between the two-dimensional spatial coordinate ${\bf r}$ and the Euclidean temporal coordinate $\tau$. The Euclidean action of our model is given by \cite{Banerjee:2019jpw},
\begin{align}
S(\psi,\psib) \ =\ -\sum_{\langle k,k'\rangle} \mathrm{Tr}(M_k M_{k'}) - \frac{U}{2} \sum_k \mathrm{Det}(M_k),
\label{eq:act}
\end{align}
where $\big(M_k\big)_{a,b} = \psi_{a,k}\psib_{b,k}$ is a $2\times 2$ matrix defined at each lattice site $k$. The symbol $\langle k,k'\rangle$ refers to neighboring sites $k$ and $k'$. The partition function is defined as usual through the Grassmann integral
\begin{align}
Z \ =\ \int \prod_{a,k}\ [\psib_{a,k}\psi_{a,k}]\ \ e^{-S(\psi,\psib)}.
\label{eq:pf}
\end{align}
It is easy to verify that the action is invariant under the $SU_L(2)\times SU_R(2)$ transformations given by $M_k \rightarrow L \ M_k\ R^\dagger$ when $k\in \mbox{even sites}$, and $M_k \rightarrow R \ M_k\ L^\dagger$, when $k\in \mbox{odd sites}$. Here we assume $L$ and $R$ are $2\times 2$ matrices, each of which are elements of the $SU(2)$ group. This means $(\psi_{1,k},\psi_{2,k})$ on even sites and $(-i\overline{\psi}_{2,k},i\overline{\psi}_{1,k})$ on odd sites transform as $SU_L(2)$ doublets, while they are singlets of $SU_R(2)$. The same fields on the opposite parity sites transform as $SU_R(2)$ doublets and $SU_L(2)$ singlets. When $U=0$, the theory has an additional $U(1)$ symmetry: $\psi_{a,x} \rightarrow e^{i\theta}\psi_{a,x}$ and $\psib_{a,x} \rightarrow e^{i\theta}\psib_{a,x}$
for the odd sites, and $\psi_{a,x} \rightarrow e^{-i\theta}\psi_{a,x}$ and $\psib_{a,x} \rightarrow e^{-i\theta}\psib_{a,x}$ on the even sites. The $U$-term, therefore, mimics the anomalous axial symmetry of the action in Quantum Chromodynamics (QCD) \cite{Cecile:2007dv}. For this reason the terms in the partition function that arise due to a non-zero value of $U$ were referred to as instantons in the earlier work. It is known that instantons are known to break the anomalous axial symmetry in QCD. In the more recent view point of qubit regularization, the instantons can be viewed as simply the local Fock vacuum states \cite{Singh:2019jog}.

\begin{figure}
\includegraphics[width=0.8\linewidth]{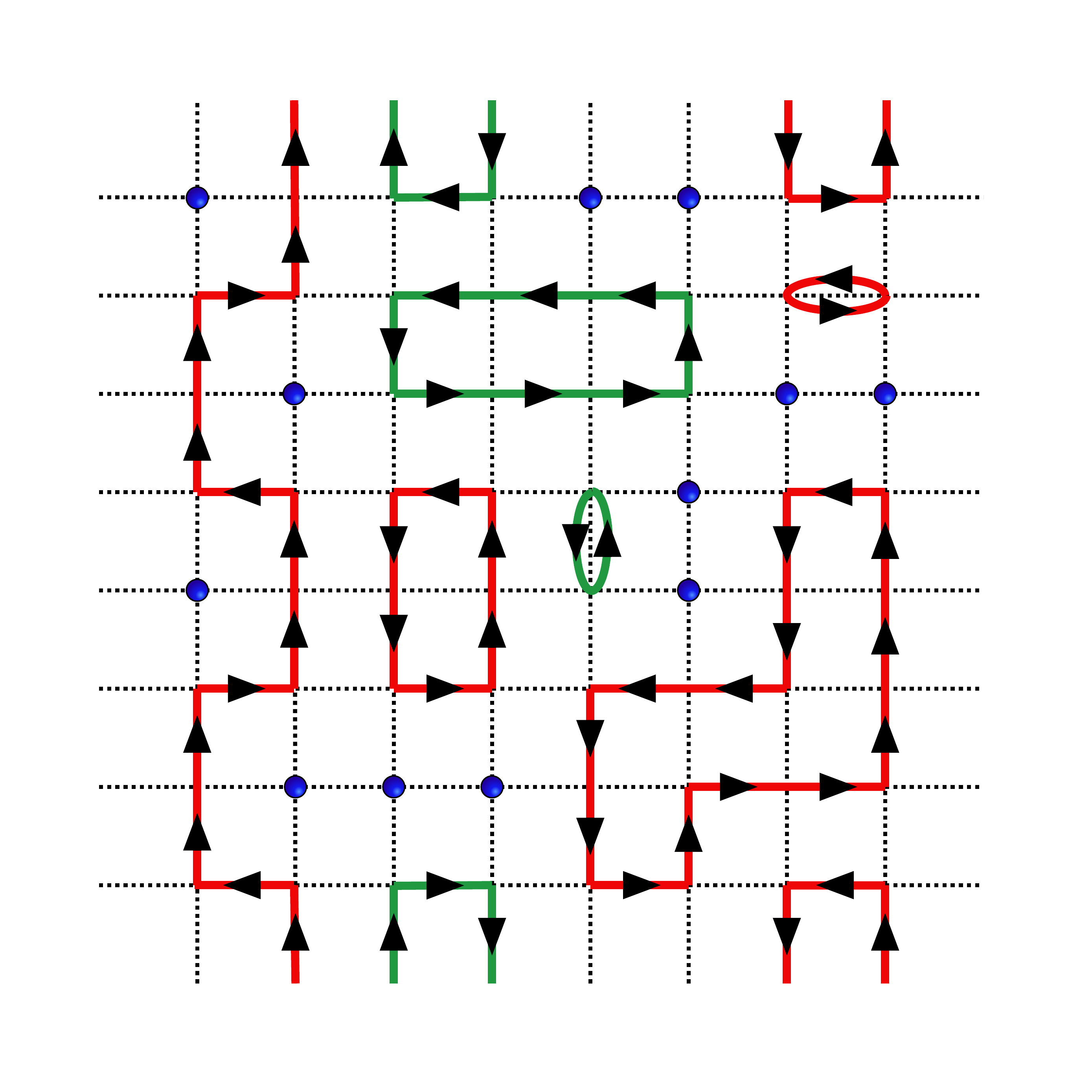}\hskip0.3in
\caption{An illustration of an $O(4)$ worldline configuration in $1+1$ dimensions with $N_I=12$ instantons (blue dots). The weight of the configuration is $U^{12}$. Each loop is an oriented loop representing a red worldline or a green worldline.}
\label{fig:wlc}
\end{figure}

It is possible to perform the Grassmann integrations in \cref{eq:pf} exactly and rewrite the partition as a sum over worldline configurations of pions and instantons \cite{Cecile:2007dv}. One then obtains
$Z\ =\ \sum_{[\ell]} U^{N_I}$, where $[\ell]$ is a configuration of worldlines, which is a collection of closed oriented loops, each of which can be in one of two colors, red or green. In addition, there are isolated sites which do not belong to the worldlines and are referred to as instantons. $N_I$ is the total number of instantons in the configuration.
An illustration of a worldline configuration on a two-dimensional lattice is shown in \cref{fig:wlc}. In an earlier work, we solved our lattice model using this worldline approach and showed that at the critical coupling of $U_c \approx 1.655394$ we can reproduce the critical scaling of the Wilson-Fisher fixed point at long distances \cite{Banerjee:2019jpw}. 

\section{Leading Charge Sectors}

Our goal is to compute the conformal dimensions $D(j_L,j_R)$ of field operators that transform in some irreducible representation $(j_L,j_R)$ of the $SU_L(2)\times SU_R(2)$ group at the critical point. We can accomplish this by computing the correlation function 
\begin{align}
C_{j_L,j_R} &= \<\overline{\cal O}_{j_L,j_R} {\cal O}_{j_L,j_R}\>\nonumber \\
&= \frac{1}{Z}\int \prod_{a,k} [\psib_{a,k}\psi_{a,k}]\ \ e^{-S(\psi,\psib)}  
\overline{\cal O}_{j_L,j_R} {\cal O}_{j_L,j_R},
\label{eq:corrfn}
\end{align}
where ${\cal O}_{j_L,j_R}$ and $\overline{\cal O}_{j_L,j_R}$ are the source and sink terms constructed with Grassmann valued fields that transform in the irreducible representation $(j_L,j_R)$. In our work, the source will be located at the temporal slice $\tau=0$, while the sink will be at the temporal slice $\tau = L/2$. Thus, at the critical point, we expect 
\begin{align}
C_{j_L,j_R} = A_{j_L,j_R} L^{-2D(j_L,j_R)} 
\label{eq:fitform}
\end{align}
for sufficiently large values of $L$. 

In order to construct 
${\cal O}_{j_L,j_R}$ and $\overline{\cal O}_{j_L,j_R}$ that transform under the irreducible representation $(j_L,j_R)$, let us
denote $\StateIn{(j_L,m_L)}{(j_R,m_R)}$ as the $(2j_L+1)(2j_R+1)$ dimensional orthonormal basis that spans the irreducible representation space of $(j_L,j_R)$. Here $-j_L \leq m_L \leq j_L$ and $-j_R \leq m_R \leq j_R$. We can label the fields that transform according to this irreducible representation as ${\cal O}_{\StateIn{(j_L,m_L)}{(j_R,m_R)}}$. For the sink terms, it is natural to choose fields that transform in the conjugate representation $\StateOut{(j_L,m_L)}{(j_R,m_R)}$. We can label these sink fields as $\overline{\cal O}_{\StateOut{(j_L,m_L)}{(j_R,m_R)}}$. While we can choose any of these fields as source and sink terms in the $(j_L,j_R)$ representation, we find that choosing $m_L=j_L$ and $m_R=j_R$ will be the most convenient choice for numerical work.

Based on the transformation property of $M_k$ it is easy to see that the four local fermion bilinear lattice fields $i\psi_{1,k}\bar{\psi}_{1,k}$, $-i\psi_{2,k}\bar{\psi}_{2,k}$, $-i\psi_{1,k}\bar{\psi}_{2,k}$, $-i\psi_{2,k}\bar{\psi}_{1,k}$
transform under the $(1/2,1/2)$ (vector) representation of $O(4)$. However, the fields transform differently on even and odd sites. The exact mapping is given in the table in \cref{tab:localf}.
\begin{table}[thb]
\centering
\setlength{\tabcolsep}{4pt}
\makegapedcells
\begin{tabular}{l|r|r}
\TopRule
\multicolumn{1}{c|}{local field} & \multicolumn{1}{c|}{even site} & 
\multicolumn{1}{c}{odd site} \\
\MidRule 
${\cal O}_{\StateIn{(1/2,1/2)}{(1/2,1/2)}}$ & 
$-i\psi_{1,k}\psib_{2,k}$ & $-i\psi_{1,k}\psib_{2,k}$ \\
${\cal O}_{\StateIn{(1/2,-1/2)}{(1/2,-1/2)}}$ &
$i\psi_{2,k}\psib_{1,k}$ & $i\psi_{2,k}\psib_{1,k}$ \\
${\cal O}_{\StateIn{(1/2,1/2)}{(1/2,-1/2)}}$ &
$i\psi_{1,k}\psib_{1,k}$ & $-i\psi_{2,k}\psib_{2,k}$ \\
${\cal O}_{\StateIn{(1/2,-1/2)}{(1/2,1/2)}}$ &
$-i\psi_{2,k}\psib_{2,k}$ & $i\psi_{1,k}\psib_{1,k}$ \\
\BotRule
\end{tabular}
\caption{\label{tab:localf} Local fermion bilinear fields transform according to the vector representation of $O(4)$. This table gives the explicit realization of each component at the lattice site $k$. As can be seen the realization can depend on whether the site is even or odd.}
\end{table}
Given the source fields we can construct the sink fields through the conjugate representation. This is given in \cref{tab:conj}.
\begin{table}[thb]
\centering
\setlength{\tabcolsep}{4pt}
\makegapedcells
\begin{tabular}{lcl}
\TopRule
\multicolumn{1}{c}{sink} 
& \multicolumn{1}{c}{} & 
\multicolumn{1}{c}{source} \\
\MidRule 
$\overline{\cal O}_{\StateOut{(1/2,1/2)}{(1/2,1/2)}}$ & = &  $-{\cal O}_{\StateIn{(1/2,-1/2)}{(1/2,-1/2)}}$ \\
$\overline{\cal O}_{\StateOut{(1/2,-1/2)}{(1/2,-1/2)}}$ & = &  $-{\cal O}_{\StateIn{(1/2,1/2)}{(1/2,1/2)}}$ \\
$\overline{\cal O}_{\StateOut{(1/2,1/2)}{(1/2,-1/2)}}$ & = &  ${\cal O}_{\StateIn{(1/2,-1/2)}{(1/2,1/2)}}$ \\
$\overline{\cal O}_{\StateOut{(1/2,-1/2)}{(1/2,1/2)}}$ & = & ${\cal O}_{\StateIn{(1/2,1/2)}{(1/2,-1/2)}}$ \\
\BotRule
\end{tabular}
\caption{\label{tab:conj} Relationship between source and sink fields on a local site.}
\end{table}
Using the information in \cref{tab:localf} and \cref{tab:conj} we can identify each worldline in \cref{fig:wlc} as a vector particle carrying an appropriate charge. For example, we can identify the sources and sinks of red worldlines as 
${\cal R}_k = -i\psi_{1,k}\psib_{2,k}$ and ${\overline{\cal R}}_k = -i\psi_{2,k}\psib_{1,k}$ on all sites. On the other hand the sources and sinks of green worldlines depend on the parity of the sites. The green source is given by ${\cal G}_k = i\psi_{1,k}\psib_{1,k}$ on even sites, and ${\cal G}_k = -i\psi_{2,k}\psib_{2,k}$ on odd sites. This is reversed for the sink of green lines. We have ${\overline{\cal G}}_k = -i\psi_{2,k}\psib_{2,k}$ on even sites, and  ${\overline{\cal G}}_k = i\psi_{1,k}\psib_{1,k}$ on odd sites. 

\begin{figure}
\includegraphics[width=0.8\linewidth]{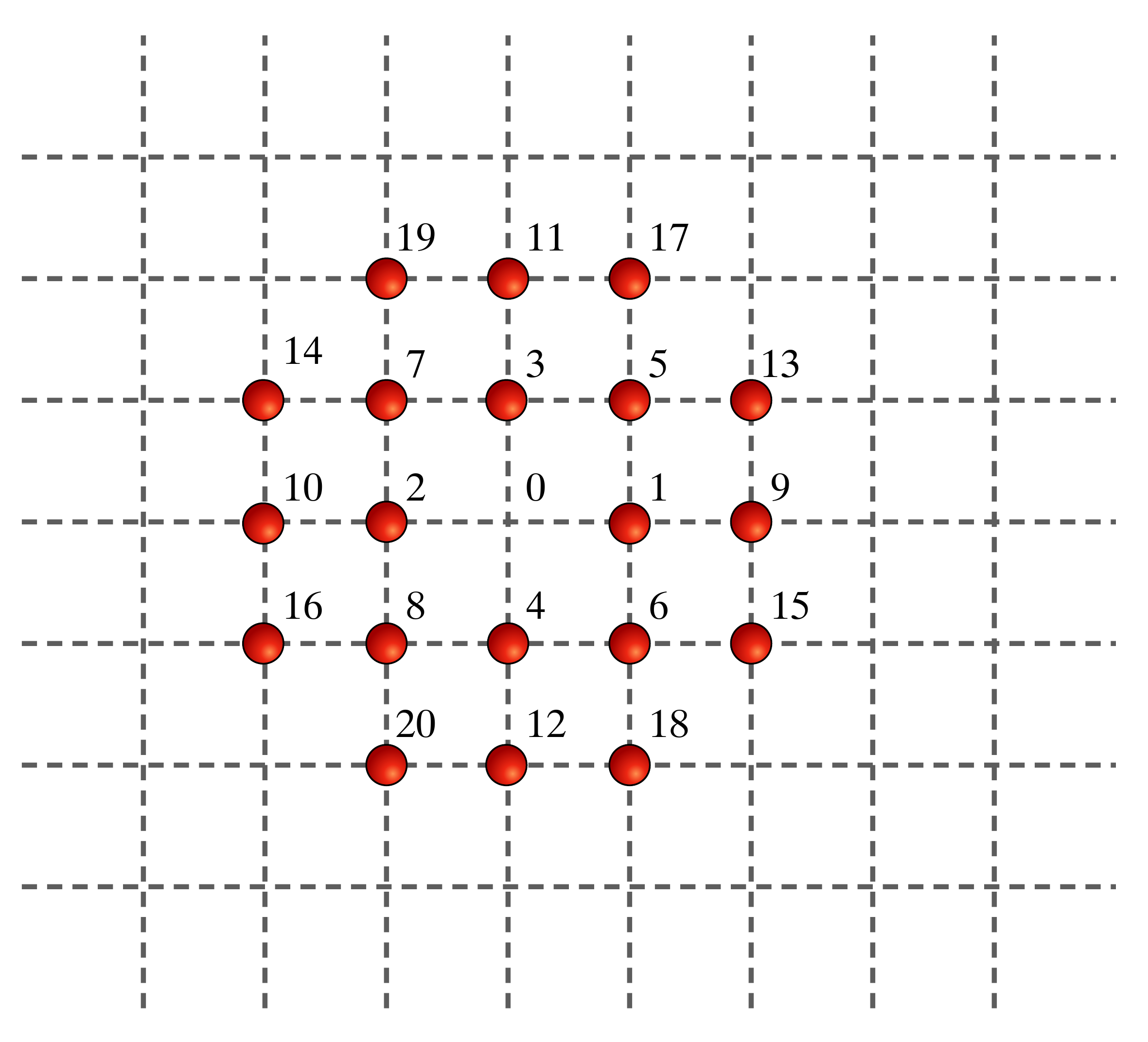}\hskip0.3in
\caption{The figure shows the arrangement of local sources of red worldlines we used to create a source in the $(j,j)$ representation. The site labeled $0$ is the origin is left empty. Depending on $j$ the sources are placed on the sites marked $k=1,2...2j$. The first $20$ sites used in our calculations up to $j=10$ are shown.}
\label{fig:src}
\end{figure}

In order to construct sources of more general irreducible representations, we use tensor product of the vector representation of
local fields distributed over several spatial lattice sites on the time slice $\tau=0$. The same sites on the time slice 
$\tau=L/2$ are used to construct the sinks. Let us first construct sources and sinks that transform in the representation $(j,j)$ which we refer to as the leading sector. We know we can construct the basis state $\StateIn{(j,j)}{(j,j)}$ as a tensor product of $2j$ states of the form $\StateIn{(1/2,1/2)}{(1/2,1/2)}$. The source ${\cal O}_{\StateIn{(1/2,1/2)}{(1/2,1/2)}}$ can also be constructed in the same way, placing $2j$ local sources of red worldlines on nearby sites on a spatial lattice. In our work we choose these lattice sites $k$ labeled as $1,2,...,2j$ around the origin as shown in \cref{fig:src}. Hence one of the possible sources in the $(j,j)$ representation is then given by ${\cal O}_{\StateIn{(j,j)}{(j,j)}}  = {\cal R}_1 {\cal R}_2...{\cal R}_{2j}$, while the corresponding sink is given by $
\overline{\cal O}_{\StateOut{(j,j)}{(j,j)}}  = \overline{\cal R}_1 \overline{\cal R}_2...\overline{\cal R}_{2j}$. Here the sites chosen for both source and sink are the same spatial sites on the $\tau=0,L/2$ time slice. These sources and sinks can be used to compute the correlation function 
\begin{align}
C_{j,j} = \big\< \overline{\cal O}_{\StateOut{(j,j)}{(j,j)}}
{\cal O}_{\StateIn{(j,j)}{(j,j)}} \big\>
\end{align}
and in order to obtain $D(j,j)$ one can use the relation $C_{j,j} \sim A_{j,j} L^{-2D(j,j)}$ for large values of $L$. In the actual worm algorithm, one in fact computes the ratio $R_j = C_{j,j}/C_{j-1,j-1}$ and fits it to the form $(A_{j,j}/A_{j-1,j-1}) L^{-2 \Delta_j}$ to compute
$\Delta_j = D(j,j)-D(j-1,j-1)$ for each value of $j$. From these differences and setting $D(0,0)=1$ one can compute $D(j,j)$. Our results from \cite{Banerjee:2019jpw} are tabulated in Table \ref{tab:CDjj} for reference.

\begin{table}
\centering
\setlength{\tabcolsep}{10pt}
\makegapedcells
\begin{tabular}{c|c||c|c}
\TopRule
$j$ & $D(j,j)$ & $j$ & D(j,j)    \\
\MidRule
1/2 & 0.515(3)  & 1 & 1.185(4)   \\
3/2 & 1.989(5)  & 2 & 2.915(6)   \\
5/2 & 3.945(6)  & 3 & 5.069(7)   \\
7/2 & 6.284(8)  & 4 & 7.575(9)   \\
9/2 & 8.949(10) & 5 & 10.386(11) \\
\BotRule
\end{tabular}
\caption{\label{tab:CDjj} Results for the conformal dimensions $D(j,j)$ up to $j=5$ computed using worldline Monte-Carlo methods 
in ~\cite{Banerjee:2019jpw}.}
\end{table}

\begin{figure}[!tbh]
\includegraphics[width=0.9\linewidth]{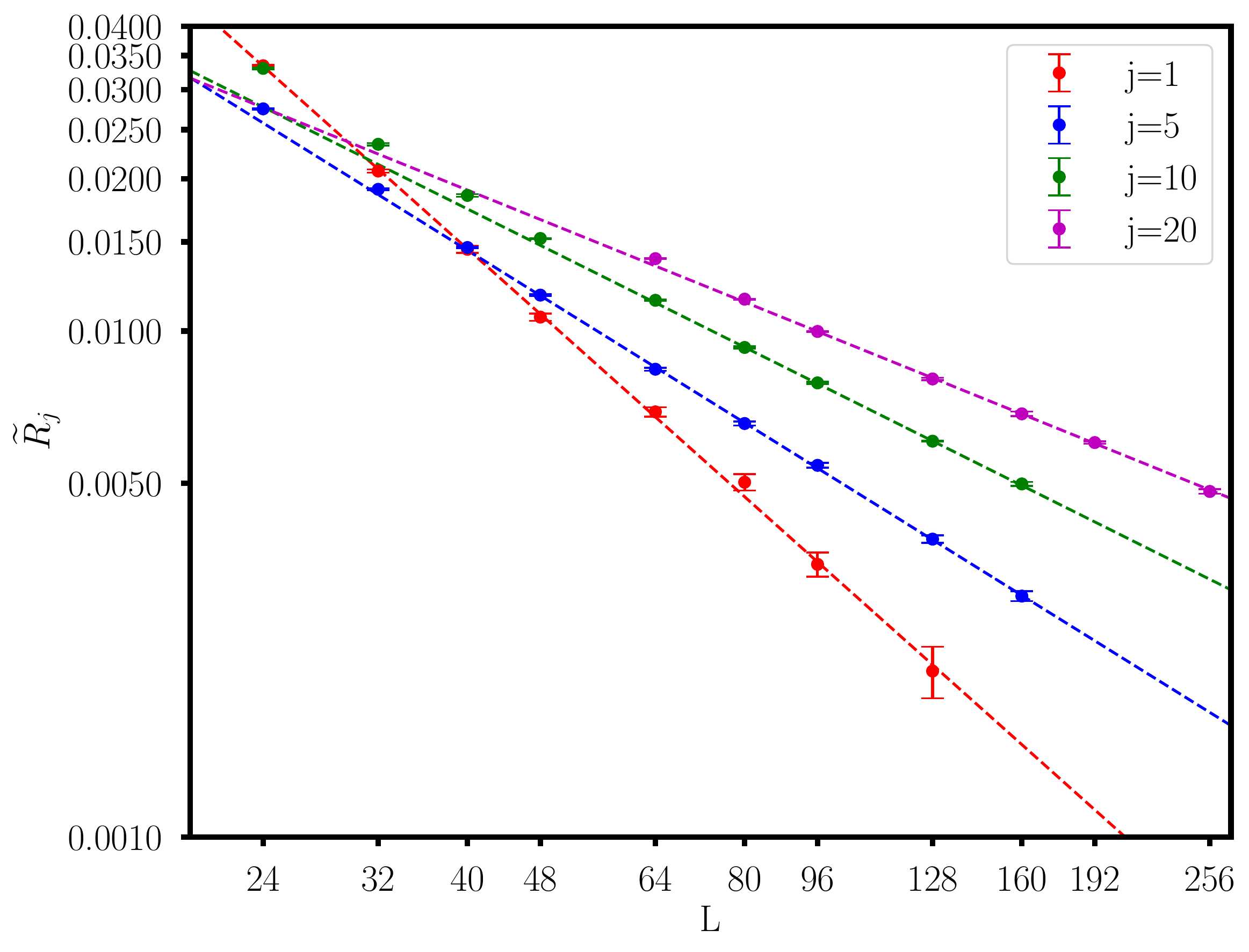}
\caption{Worldline Monte Carlo results for the ratio $\tilde{R}_j$ as a function of $L$ for various values of $j$. The solid lines show the fits given in \cref{tab:res1}.}
\label{fig:powfits}
\end{figure}

\section{Subleading Sector}

In this work we extend our earlier results in the leading sector and compute the conformal dimensions of the subleading sector, $D(j,j-1)$ for $j \geq 1$. For this we need to construct source and sink operators that transform in the representation $(j,j-1)$. We know we can construct the state $\StateIn{j, j}{j-1, j-1}$ by applying the lowering operator $J_R^{-}$ to the $\StateIn{j,j}{j,j}$ and then constructing orthogonal states in the tensor product space. This procedure naturally leads to $2j-1$ orthonormal states, which we can label with an additional index $M=1,2...,(2j-1)$. Translating this to the construction of sources, we now introduce the source ${\cal O}_\ell = {\cal R}_1 {\cal R}_2...{\cal G}_\ell...{\cal R}_{2j}$ where the red source on one lattice site $\ell$ is replaced by a green source where $\ell=1,2...,2j$. Similarly, we introduce the corresponding sinks as
$\overline{\cal O}_\ell = \overline{\cal R}_1 \overline{\cal R}_2...\overline{\cal G}_\ell...\overline{\cal R}_{2j}$. We can then argue that 
\begin{align}
{\cal O}_{\StateIn{(j,j)}{(j,j-1)}}  & = 
\frac{1}{\sqrt{2j}}\sum_{\ell=1}^{2j} {\cal O}_\ell.
\label{eq:ojj1}
\end{align}
where the right-hand side is a sum over the $2j$ source terms we introduced above. Note that there are $2j-1$ sources orthogonal to \cref{eq:ojj1}, which can label with $M=1,2...,2j-1$ as before. These will naturally transform in the $(j,j-1)$ representation. Explicitly these sources are given by
\begin{align}
{\cal O}^M_{\StateIn{(j,j)}{(j-1,j-1)}}  & = 
\frac{1}{\sqrt{2j}}\sum_{\ell=1}^{2j} e^{i2\pi (\ell-1) M/(2j)} {\cal O}_\ell.
\label{eq:ojj-1src}
\end{align}
We can similarly define the corresponding $2j-1$ sinks as
\begin{align}
\overline{\cal O}^M_{\StateOut{(j,j)}{(j-1,j-1)}}  & = \frac{1}{\sqrt{2j}}\sum_{\ell=1}^{2j} e^{-i2\pi (\ell-1) M/(2j)} \overline{\cal O}_\ell.
\label{eq:ojj-1snk}
\end{align}
Since all $2j-1$ sources and sinks transform under the same irreducible representation $(j,j-1)$ any combination of them can be used in \cref{eq:corrfn} to extract $D(j,j-1)$. Let us define the correlation matrix
\begin{align}
C^{MM'}_{j,j-1} = \big\< 
\overline{\cal O}^M_{\StateOut{(j,j)}{(j-1,j-1)}} \ {\cal O}^{M'}_{\StateIn{(j,j)}{(j-1,j-1)}}\big\>
\label{eq:corrfn-jj-1}
\end{align}
where we expect $C^{M M'}_{j,j-1} \sim A^{MM'}_{j,j-1} L^{-2D(j,j-1)}$ for large values of $L$. Practically it is more convenient to compute the average of the trace of this correlation matrix 
which can be simplified to the form
\begin{align}
C_{j,j-1} \ =\ \sum_\ell 
\Big\{ \frac{1}{2j}
\big\<\overline{\cal O}_\ell {\cal O}_\ell\big\> - \frac{1}{2j(2j-1)}
\sum_{\ell' \neq \ell} 
\big\<\overline{\cal O}_\ell {\cal O}_{\ell'}\big\>
\Big\}.
\label{eq:obs}
\end{align}
Note that we also expect $C_{j,j-1} \sim A_{j,j-1} L^{-2D(j,j-1)}$.

\begin{table}[b]
\centering
\setlength{\tabcolsep}{4pt}
\makegapedcells
\begin{tabular}{r|c|c|c|c}
\TopRule
\multicolumn{1}{c|}{$j$} & \multicolumn{1}{c|}{$L$-range} & \multicolumn{1}{c|}{$A_{j,j-1}/A_{j,j}$} &  \multicolumn{1}{c|}{$\tD(j)$} & \multicolumn{1}{c}{$\chi^2$/DOF}\\
\MidRule 
$1$    & $24-128$ &  $5.87(25)$ & $0.813(6)$  & $1.11$ \\
$3/2$  & $24-128$ &  $2.50(11)$ & $0.750(6)$  & $0.62$ \\
$2$    & $24-96$  &  $2.13(06)$ & $0.722(4)$  & $0.26$ \\
$5/2$  & $32-96$  &  $1.75(08)$ & $0.685(6)$  & $1.28$ \\
$3$    & $32-96$  &  $1.54(08)$ & $0.659(7)$  & $0.93$ \\
$7/2$  & $32-96$  &  $1.35(05)$ & $0.633(5)$  & $0.38$ \\
$4$    & $32-96$  &  $1.18(04)$ & $0.607(4)$  & $0.40$ \\
$9/2$  & $40-160$ &  $1.05(04)$ & $0.586(5)$  & $0.94$ \\
$5$    & $40-160$ &  $0.94(04)$ & $0.566(5)$  & $0.89$ \\
$11/2$ & $48-160$ &  $0.90(03)$ & $0.555(4)$  & $0.66$ \\
$6$    & $48-160$ &  $0.83(03)$ & $0.541(5)$  & $1.40$ \\
$13/2$ & $64-160$ &  $0.75(04)$ & $0.525(7)$  & $1.11$ \\
$7$    & $64-160$ &  $0.71(03)$ & $0.513(5)$  & $1.18$ \\
$15/2$ & $64-160$ &  $0.69(04)$ & $0.506(6)$  & $1.45$ \\
$8$    & $64-160$ &  $0.60(03)$ & $0.486(5)$  & $0.77$ \\
$17/2$ & $64-160$ &  $0.61(03)$ & $0.484(5)$  & $0.83$ \\
$9$    & $80-160$ &  $0.53(04)$ & $0.467(8)$  & $0.98$ \\
$19/2$ & $80-160$ &  $0.53(03)$ & $0.463(7)$  & $0.47$ \\
$10$   & $80-160$ &  $0.50(02)$ & $0.454(5)$  & $0.63$ \\
$20$   & $96-256$ &  $0.28(01)$ & $0.367(3)$  & $0.65$ \\
\BotRule
\end{tabular}
\caption{\label{tab:res1} Results of the fit of $\tilde{R}_j$ shown in \cref{fig:powfits} to the form $(A_j/A_{j-1})L^{-2\tilde{\Delta}_j}$. The range of $L$ values used in the fit are given in the second column. We observe that as $j$ increases this range needs to involve larger lattice sizes for a good fit.}
\end{table}

As in the leading sector, in the worldline algorithm it much easier to compute the ratio $\tilde{R}_j = C_{j,j-1}/C_{j,j}$. For this one constructs a worldline Monte Carlo method to generate configurations with $2j$ red sources and sinks that contribute to $C_{j,j}$. In every configuration of this ensemble, we then imagine flipping each of the $2j$ sources located at the sites $\ell=1,2...,2j$ to a green source. The worldline of the green source then naturally travels through the lattice to a sink at some location $\ell'$. Then we compute the contribution to $\tilde{R}_j$ from that configuration using \cref{eq:obs}, which means we add $1/2j$ if $\ell = \ell'$, or subtract the value $1/(2j(2j-1))$ if $\ell \neq \ell'$ for every value of $\ell$. Averaging this contribution over the ensemble of configurations generated by the worldline algorithm gives us the ratio, $\tilde{R}_j$ which is expected to scale as $(A_j/A_{j-1})L^{-2\tilde{\Delta}_j}$ where $\tilde{\Delta}_j = D(j,j-1)-D(j,j)$. Using the values of $D(j,j)$ we compute $D(j,j-1)$.

\begin{table}[h]
\setlength{\tabcolsep}{4pt}
\makegapedcells
\centering
\begin{tabular}{c|c|c|c|c|c}
\TopRule
 $j$ & $\lambda_0$  & $\lambda_{1/2}$  & $\lambda_1$ & $\lambda_{3/2}$ & $\chi^2$ \\
 range & & & & & /DOF \\
\MidRule
1-20 & 0.07(2) & 1.5(1) & -1.0(2) & 0.3(1) & 0.6 \\
1-20 & 0 & 1.42(1) & 0 & -0.69(1) & 48 \\
1-20 & 0 & 1.96(2) & -1.83(6) & 0.69(5) & 1.8 \\
1-20 & 0.16(1) & 0.98(2) & 0 & -0.34(2) & 2.3 \\  
1.5-20 & 0.13(1) & 1.07(2) & 0 & -0.47(3) & 0.3 \\
2-20 & 0 & 2.06(3) & -2.27(14) & 1.14(13) & 1.1 \\
8-20 & 0 & 1.81(2) & 0 & -3.5(2) & 1.1 \\
8-20 & 0 & 2.09(4) & -2.02(13) & 0 & 0.6 \\
\BotRule
\end{tabular}
\caption{\label{tab:res2} Fits of $\tD_j$ to the functional form given in \cref{eq:lqpred}. We first consider the whole range of $j$ in the first four rows. While including all four coefficients as fitting parameters gives an excellent fit, setting the two purely quantum mechanical terms $\lambda_0=\lambda_1=0$ makes the fit quite bad. Including even one of them is sufficient to improve the fit considerably. Setting two of the fitting parameters to zero is only possible by shrinking the allowed region of $j$ considerably.}
\end{table}

\section{Results}
We have performed a series of Monte Carlo calculations for $1 \leq j \leq 10$ in increments of $1/2$, and for $j=20$. For runs till $j=10$, the lattice sizes ranged from $L=32$ up to $L=160$. For each value of $j$ we have always found that our data fits well to the expected form $C_{j,j-1}/C_{j,j} \sim L^{-2\tilde{\Delta}_j}$ for sufficiently large values of $L$. As examples, fits for $j=1,5$, $10$, and $20$ are illustrated in \cref{fig:powfits}. From the figure, we observe that as $j$ increases the range of lattice sizes where a simple power law emerges changes to larger $L$ values. The range of $L$ values where we perform the fits and the obtained fit parameters for various values of $j$ are tabulated in \ref{tab:res1}. At $j=20$ we have extended our calculations up to lattice sizes of $L=256$.

\begin{figure}[t]
\includegraphics[width=0.9\linewidth]{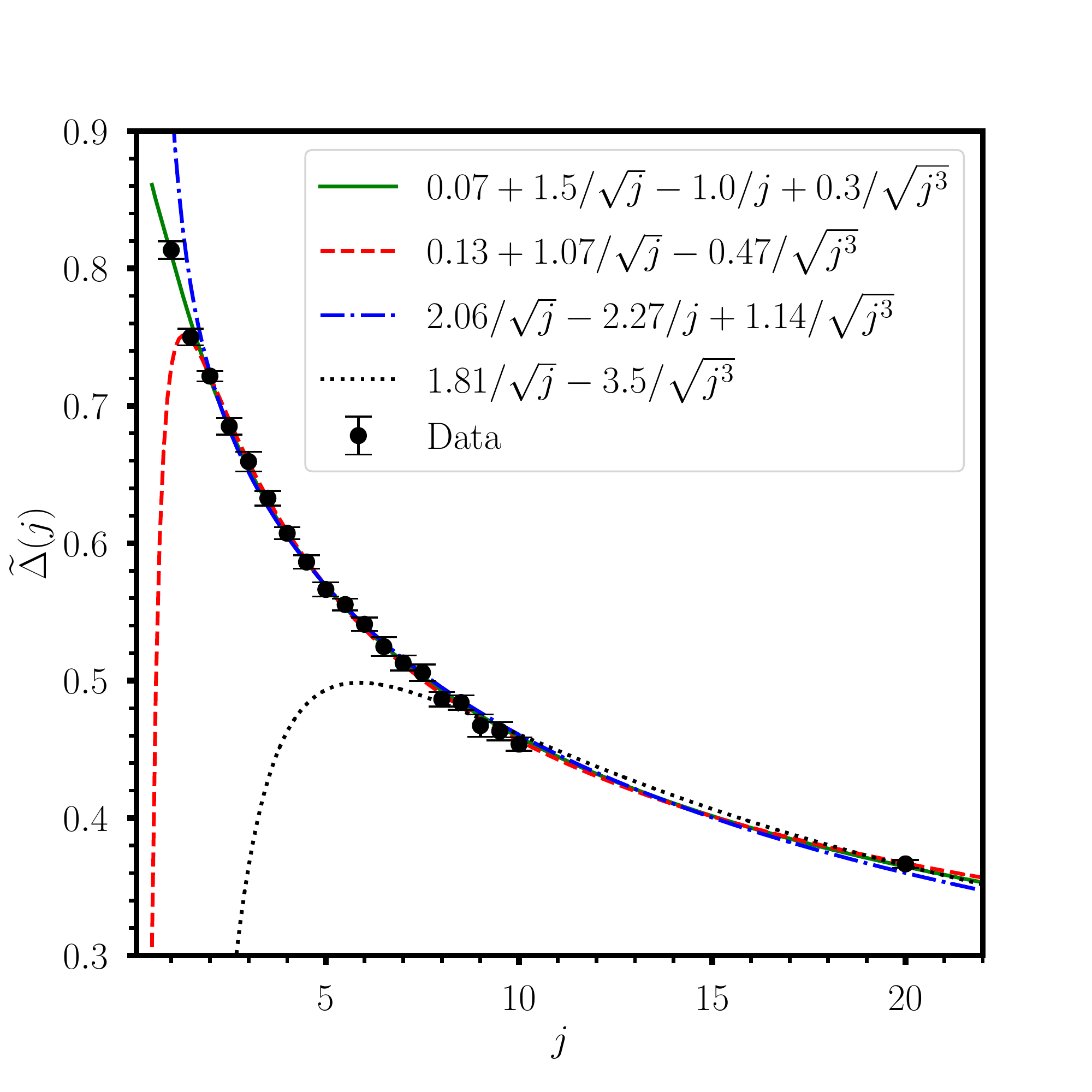}\hskip0.3in
\caption{The plot of $\tD_j$ as a function of $j$ from  \cref{tab:res1}. The lines shown are fits to the form given in \cref{eq:lqpred} with parameter values given in the first (solid), fifth (dashes), sixth (dashes+dots) rows and seventh (dots) rows of  \cref{tab:res2}.}
\label{fig:delta}
\end{figure}

Having obtained the values of $\tilde{\Delta}_j$ for various values of $j$ we try to extract the constants $\lambda_i$'s in the large charge expansion based on \cref{eq:lqpred}, assuming we can neglect ${\cal O}(1/j^2)$ terms. We try several fits to understand the role of the purely quantum terms $\lambda_0$ and $\lambda_1$. These fits are shown in \cref{tab:res2}. First, we note that our data for the entire range of $j$ values fits well to the form \cref{eq:lqpred}, if we assume all four coefficients are non-zero (first row in \cref{tab:res2}). On the other hand if we drop both quantum terms the fit becomes quite bad (second row in \cref{tab:res2}). The presence of either of the two quantum terms is sufficient to bring down the $\chi^2/DOF$ considerably. For example setting $\lambda_1=0$, we can get an excellent fit if we just drop the $j=1$ data (fifth row in \cref{tab:res2}). On the other hand setting $\lambda=0$ and dropping both $j=1,1.5$ makes the fit acceptable (sixth row in \cref{tab:res2}). If we drop both quantum terms (i.e., set $\lambda_0=\lambda_1=0$) we can only get a good fit in the range $j=8-20$. We believe this is just an artifact of a small range of $j$ as can be seen in \cref{fig:delta}.

Clearly, our data is consistent with the conjecture that $\lambda_0 = 0$ (see sixth row in \cref{tab:res2}). Assuming this, we can try to determine the leading two terms more reliably by fitting our data under the constraint $\lambda_{3/2}=0$. In this case we can get a good fit only in the smaller range $j=8-20$ (eighth row in \cref{tab:res2}). As expected, this changes $\lambda_{1/2}$ and $\lambda_1$ slightly. Taking such systematic fitting effects into account, we estimate that $\lambda_{1/2}=2.08(5)$, $\lambda_1=2.2(3)$. Unfortunately, calculations at higher values of $j$ are difficult since we need larger lattice sizes, but they can be obtained with more computing resources and can help confirm the conjecture that $\lambda_0=0$.

\section{Conclusions}

In this work we have constructed a Monte Carlo method to compute the sub-leading conformal dimensions in the large charge expansion at the $O(4)$ Wilson-Fisher fixed point. We used this method to compute $\tD_j$ for several values of $j$ in the range $1 \leq j\leq 20$. While our results are consistent with the general predictions of the expansion given in \cref{eq:lqpred}, we cannot rule out the possibility that $\tD_j$ approaches a non-zero constant $\lambda_0$ in the large $j$ limit. However, our data is consistent with the conjecture that  $\lambda_0=0$. Assuming this to be true, we can estimate the leading two terms in the expansion to be $\lambda_{1/2}=2.08(5)$, $\lambda_1=2.2(3)$. Since $\lambda_1$ must be a calculable number within the large charge effective field theory, we hope our work will motivate someone to calculate it in the future. Calculations at large $j$ may also help determine $\lambda_0$ reliably.

\section{Acknowledgments}

We thank D.~Orlando, S.~Reffert for extensive discussions and collaborating with us in the past. We also thank S.~Hellerman for useful discussions about this work. The material presented here is based upon work supported by the U.S. Department of Energy, Office of Science, Nuclear Physics program under Award Number DE-FG02-05ER41368. 

\bibliographystyle{apsrev4-1}
\bibliography{refs}
\end{document}